\documentclass[11pt,twoside]{article}

\usepackage{asp2014}

\aspSuppressVolSlug
\resetcounters

\begin{document}

\title{Quantum Convolutional Neural Networks for the detection of Gamma-Ray Bursts in the AGILE space mission data}

\author{A. Rizzo,$^1$ N. Parmiggiani,$^2$, A. Bulgarelli $^2$, A. Macaluso$^3$, V. Fioretti$^2$, L. Castaldini$^2$, A. Di Piano$^{4,2}$, G. Panebianco$^{5,2}$, C. Pittori$^{6,7}$, M. Tavani$^8$, C. Sartori$^9$, C. Burigana$^{10}$, V. Cardone$^6$, F. Farsian$^{11}$, M. Meneghetti$^2$, G. Murante$^{12}$, R. Scaramella$^6$, F. Schillir\`{o}$^{11}$, V. Testa$^6$ and T. Trombetti$^{10}$}
\affil{$^1$ University of Bologna, Department of Computer Science and Engineering (DISI), Viale del Risorgimento 2, 40136, Bologna, Italy; \email{alessandro.rizzo14@studio.unibo.it}}
\affil{$^2$ INAF/OAS Bologna, Via P. Gobetti 93/3, 40129 Bologna, Italy}
\affil{$^3$ German Research Center for Artificial Intelligence (DFKI), 66123 Saarbruecken, Germany.}
\affil{$^4$ Universit\`{a} degli Studi di Modena e Reggio Emilia, DIEF, Via Pietro Vivarelli 10, 41125, Modena, Italy}
\affil{$^5$ Department of Physics and Astronomy, University of Bologna, Via Gobetti 93/2, 40129, Bologna, Italy}
\affil{$^6$ INAF/OA Roma, Via Frascati 33, I-00078 Monte Porzio Catone, Roma, Italy.}
\affil{$^7$ ASI/SSDC Roma, Via del Plitecnico snc, I-00133, Roma, Italy.}
\affil{$^8$ INAF/IAPS Roma, Via del Fosso del Cavaliere 100, 00133, Roma, Italy.}
\affil{$^9$ University of Bologna, Department of Computer Science and Engineering (DISI), Via dell'Universit\`{a} 50, I-4572 Cesena, Italy}
\affil{$^{10}$ INAF/IRA Bologna, Via P. Gobetti, 101, 40129 Bologna, Italy}
\affil{$^{11}$ INAF/OA Catania, Via S. Sofia 78, 95123, Catania, Italy}
\affil{$^{12}$ INAF/OA Trieste, Via Giambattista Tiepolo 11, 34131, Trieste, Italy}

\paperauthor{Alessandro~Rizzo}{alessandro.rizzo14@studio.unibo.it}{0009-0003-4341-2988}{University of Bologna}{Department of Computer Science and Engineering (DISI)}{Bologna}{}{40136}{Italy}
\paperauthor{Nicol\`{o}~Parmiggiani}{nicolo.parmiggiani@inaf.it}{0000-0002-4535-5329}{INAF}{OAS}{Bologna}{BO}{40129}{Italy}
\paperauthor{Andrea~Bulgarelli}{andrea.bulgarelli@inaf.it}{0000-0001-6347-0649}{INAF}{OAS}{Bologna}{BO}{40129}{Italy}
\paperauthor{Antonio~Macaluso}{antonio.macaluso.90@gmail.com }{0000-0002-1348-250X}{}{DFKI}{Saarbruecken}{66123}{}{Germany}
\paperauthor{Valentina~Fioretti}{valentina.fioretti@inaf.it}{0000-0002-6082-5384}{INAF}{OAS}{Bologna}{BO}{40129}{Italy}
\paperauthor{Luca~Castaldini}{luca.castaldini@inaf.it}{0009-0000-5501-4328}{INAF}{OAS}{Bologna}{BO}{40129}{Italy}
\paperauthor{Ambra~Di~Piano}{ambra.dipiano@inaf.it}{0000-0002-9894-7491}{INAF}{OAS}{Bologna}{BO}{40129}{Italy}
\paperauthor{Gabriele~Panebianco}{gabriele.panebianco@inaf.it}{0000-0002-3410-8613}{INAF}{OAS}{Bologna}{BO}{40129}{Italy}
\paperauthor{Carlotta~Pittori}{carlotta.pittori@inaf.it}{0000-0001-6661-9779}{INAF}{OAR}{Monte Porzio Catone}{RO}{00078}{Italy}
\paperauthor{Marco~Tavani}{marco.tavani@inaf.it}{0000-0003-2893-1459}{INAF}{OAR}{Monte Porzio Catone}{RM}{00078}{Italy}
\paperauthor{Claudio~Sartori}{claudio.sartori@unibo.it}{0000-0003-4535-1026}{University of Bologna}{Deparment of Computer Science and Engineering (DISI)}{Forli-Cesena}{FC}{I-47522}{Italy}
\paperauthor{Carlo~Burigana}{carlo.burigana@inaf.it}{0000-0002-3005-5796}{INAF}{IRA}{Bologna}{BO}{40129}{Italy}
\paperauthor{Vincenzo~Cardone}{vincenzo.cardone@inaf.it}{0000-0002-2079-7438}{INAF}{OAS}{Bologna}{BO}{40129}{Italy}
\paperauthor{Farida~Farsian}{farian.farsian@inaf.it}{}{INAF}{OA}{Catania}{CT}{95123}{Italy}
\paperauthor{Massimo~Meneghetti}{massimo.meneghetti@inaf.it}{0000-0003-1225-7084}{INAF}{OAS}{Bologna}{BO}{40129}{Italy}
\paperauthor{Giuseppe~Murante}{giuseppe.murante@inaf.it}{0000-0002-5155-130X}{INAF}{OA}{Trieste}{TS}{34131}{Italy}
\paperauthor{Roberto~Scaramella}{roberto.scaramella@inaf.it}{0000-0003-2229-193X}{INAF}{OAR}{Monte Porzio Catone}{RM}{00078}{Italy}
\paperauthor{Francesco~Schillir\`{o}}{francesco.schilliro@inaf.it}{0000-0001-5106-2277}{INAF}{OA}{Catania}{CT}{95123}{Italy}
\paperauthor{Vincenzo~Testa}{vincenzo.testa@inaf.it}{0000-0003-1033-1340}{INAF}{OAR}{Monte Porzio Catone}{RM}{00078}{Italy}
\paperauthor{Tiziana~Trombetti}{tiziana.trombetti@inaf.it}{0000-0001-5166-2467}{INAF}{IRA}{Bologna}{BO}{40129}{Italy}



\begin{abstract}
Quantum computing represents a cutting-edge frontier in artificial intelligence. It makes use of hybrid quantum-classical computation which tries to leverage quantum mechanic principles that allow us to use a different approach to deep learning classification problems. The work presented here falls within the context of the
AGILE space mission, launched in 2007 by the Italian Space Agency. We implement different Quantum Convolutional Neural Networks (QCNN) that analyze data acquired by the
instruments onboard AGILE to detect Gamma-Ray Bursts from sky maps or light curves. We use
several frameworks such as TensorFlow-Quantum, Qiskit and PennyLane to simulate a quantum computer. We achieved an accuracy of 95.1\% on sky maps with QCNNs, while the classical counterpart achieved 98.8\% on the same data, using however hundreds of thousands more parameters.
\end{abstract}



\section{Introduction}
\label{sec:Intro}
AGILE, \citep{tavani2008}, is a space mission launched in 2007 by the Italian Space Agency. The AGILE payload consists of the Silicon Tracker (ST), the SuperAGILE X-ray detector, the Mini-Calorimeter (MCAL) and an AntiCoincidence System (ACS). The combination of those elements composes the Gamma-Ray Imaging Detector (GRID) \citep{BULGARELLI2010213}. The latter is used to detect different transient astronomical events, such as Gamma-Ray Bursts (GRBs). The goal of this work is to test several Quantum Convolutional Neural Networks (QCNN), the quantum counterpart of a classical Convolutional Neural Network, in order to detect GRBs from sky maps and time series generated by the AGILE data called light curves. In this work, sky maps are considered as intensity maps which are count maps divided by the exposure and they are produced by the GRID instrument; the dataset of sky maps has been simulated using an AGILE software which creates both new GRB and background maps \citep{Parmiggiani_2021}. Light curves show the number of photon counts detected during time and binned in time windows with the same size. We analyzed the light curves obtained with the ratemeters of the ACS \citep{Parmiggiani_2023} and compared the results with those obtained with classical deep learning.\newline
We used quantum computing techniques to analyze these data. A quantum machine works on quantum bits, called qubits. Those can exist in multiple states simultaneously, known as superposition, and can be entangled. Those properties can bring an advantage over the classical approach. The fundamental building block of a quantum deep learning architecture is a parameterized quantum circuit, also known as ansatz. We implement a quantum data encoding layer to efficiently represent and input classical data into a quantum system.\newline

\articlefigure[width=.6\textwidth]{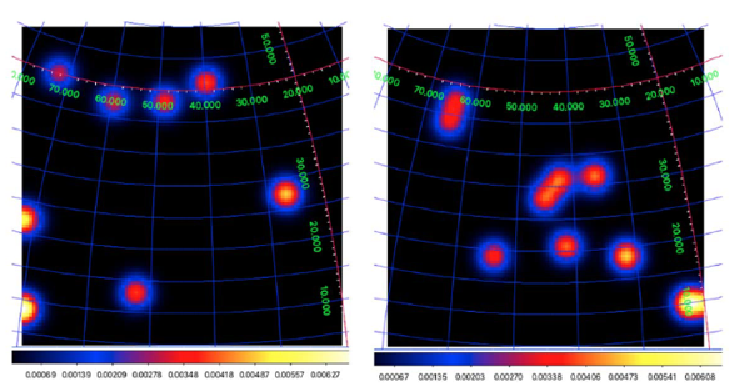}{maps}{Example of sky maps representing background noise (on the left) and a GRB signal (on the right).}

\articlefigure[width=.7\linewidth]{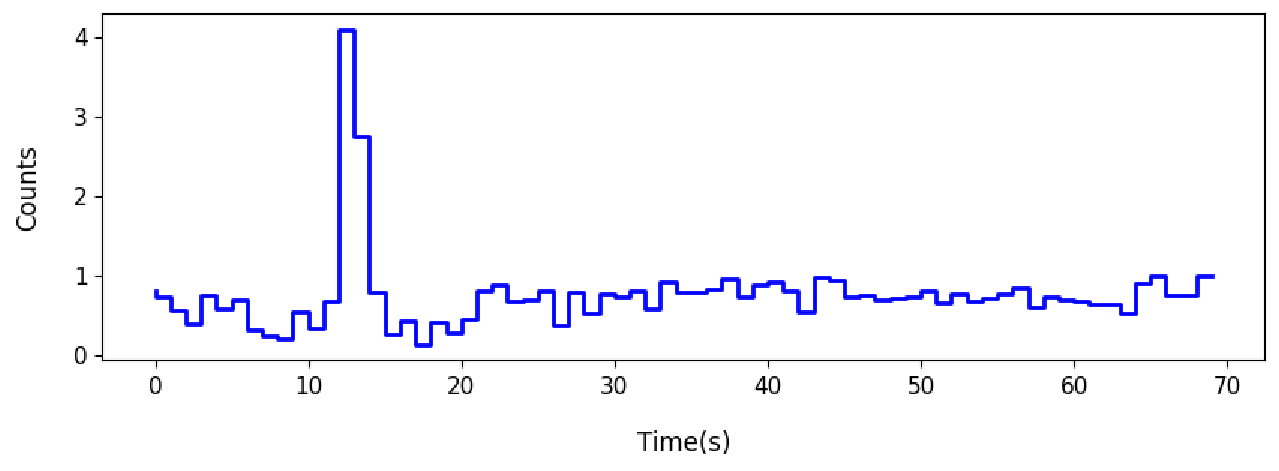}{lc}{Example of a rebinned light curve representing the signal GRB100719D. The x-axis represents the time window with bins of 2.048 seconds. On the y-axis, the count rate of photons detected by the ACS over time.}

\section{Methodology}
The two datasets used to train the models are composed of two classes: GRB and background noise. Sky maps are split between 400 GRBs and 400 background maps and each image is of size $33\times33 $ pixels with a diameter of $50^{\circ}$ and with each pixel being $1.5^{\circ}$. We had to reduce the original size of sky maps, which was $100 \times 100$, in order to ease the training process of the quantum model. We also applied Gaussian smoothing to represent the point spread function of the AGILE satellite. \newline
Regarding light curves, we used a dataset of univariate time series composed of 43 GRBs samples and 101 background samples. We applied rebinning and normalization to reduce the size of the data to use them as input to a QNN. \newline
We used different types of encoding such as angle embedding, amplitude embedding and data re-uploading \citep{P_rez_Salinas_2020} to encode the classical data into quantum states. We tested all of them with sky maps, but only the latter with light curves. We trained solely the parameters that compose the quantum convolutional circuit. We tested different QCNNs using Tensorflow-Quantum \citep{broughton2021tensorflow}, Qiskit \citep{Qiskit} and PennyLane \citep{bergholm2022pennylane}. We tried both fully quantum and hybrid classical-quantum architectures.

\section{Results}
We tested each QCNN using a training set composed of either sky maps or light curves and different embedding techniques. In the top part of Table \ref{tab:results} the reader can see the best results obtained with sky maps. The best performing ansatz for the fully quantum network was taken from \citep{Hur_2022}, implemented over a QCNN composed of 8 qubits. 
The results we achieved with light curves tend to be worse as it is shown in the bottom part of Table \ref{tab:results}. This might be due to the limitation of the very small dataset of time series available for this work. As with sky maps, the fully quantum network which reached the best results is composed of 8 qubits.   

\begin{table}[!h]
\resizebox{\linewidth}{!}{\begin{tabular}{|llllll|}
\hline
\multicolumn{1}{|c|}{\textbf{Approach}} &
  \multicolumn{1}{c|}{\textbf{\begin{tabular}[c]{@{}c@{}}Best train \\ accuracy\end{tabular}}} &
  \multicolumn{1}{c|}{\textbf{\begin{tabular}[c]{@{}c@{}}Best validation \\ accuracy\end{tabular}}} &
  \multicolumn{1}{c|}{\textbf{\begin{tabular}[c]{@{}c@{}}Training \\ time\end{tabular}}} &
  \multicolumn{1}{c|}{\textbf{\begin{tabular}[c]{@{}c@{}}Number of \\ epochs\end{tabular}}} &
  \multicolumn{1}{c|}{\textbf{Parameters}} \\ \hline
\multicolumn{6}{|c|}{\textbf{Sky Maps}}                                                                                                                                 \\ \hline
\multicolumn{1}{|c|}{Classical}     & \multicolumn{1}{c|}{0.977} & \multicolumn{1}{c|}{0.988} & \multicolumn{1}{c|}{4.2 s}       & \multicolumn{1}{c|}{12}  & \multicolumn{1}{c|}{1,072,000} \\ \hline
\multicolumn{1}{|c|}{Hybrid}        & \multicolumn{1}{c|}{0.976} & \multicolumn{1}{c|}{0.981} & \multicolumn{1}{c|}{8 min 15 s}  & \multicolumn{1}{c|}{53}  & \multicolumn{1}{c|}{288}       \\ \hline
\multicolumn{1}{|c|}{Fully Quantum} & \multicolumn{1}{c|}{0.945} & \multicolumn{1}{c|}{0.951} & \multicolumn{1}{c|}{25 min 50 s} & \multicolumn{1}{c|}{101} & \multicolumn{1}{c|}{51}        \\ \hline
\multicolumn{6}{|c|}{\textbf{Light Curves}}                                                                                                                             \\ \hline
\multicolumn{1}{|c|}{Classical}     & \multicolumn{1}{c|}{0.990} & \multicolumn{1}{c|}{0.955} & \multicolumn{1}{c|}{10.8 s}      & \multicolumn{1}{c|}{91}  & \multicolumn{1}{c|}{2,680}     \\ \hline
\multicolumn{1}{|c|}{Hybrid}        & \multicolumn{1}{c|}{0.988} & \multicolumn{1}{c|}{0.828} & \multicolumn{1}{c|}{1 min 58 s}  & \multicolumn{1}{c|}{59}  & \multicolumn{1}{c|}{580}       \\ \hline
\multicolumn{1}{|c|}{Fully Quantum} & \multicolumn{1}{c|}{0.862} & \multicolumn{1}{c|}{0.810} & \multicolumn{1}{c|}{8 min 39 s}  & \multicolumn{1}{c|}{200} & \multicolumn{1}{c|}{16}        \\ \hline
\end{tabular}}
\caption{Results obtained with sky maps (above) and light curves (below).}
\label{tab:results}
\end{table}

\section{Conclusions \& Future Works}
We developed several QCNN architectures to detect GRBs from the sky maps and light curves obtained with the AGILE detectors. We evaluated the trained models on a separated test set obtaining promising results. We managed to achieve a good accuracy with QCNNs, very close to the classical models, using anyway fewer parameters. However, the training times are longer with the quantum approach; in fact we used a simulator instead of a real quantum machine, for this reason we plan to test the QCNNs on a real quantum computer to see if the training times can be sped up. Moreover, quantum deep learning is still in its early stages, and optimization and training algorithms are still not very well suited for a fully quantum approach. It is also possible to further explore the possibilities offered by quantum deep learning to analyze the data captured from other AGILE detectors, this is left for future work. Finally, the knowledge acquired in this project can be used in the next generation of high-energy facilities such as CTA \citep{2018} or COSI \citep{tomsick2019compton}.\newline

\acknowledgements
The AGILE Mission is funded by the Italian Space Agency (ASI) with scientific and programmatic participation by the Italian National Institute for Astrophysics (INAF) and the Italian National Institute for Nuclear Physics (INFN). This investigation is supported by the ASI grant I/028/12.7-2022 and also partially funded with an INAF "Mini-Grant" 2022. We thank the ASI management for unfailing support during AGILE operations. We acknowledge the effort of ASI and industry personnel in operating the  ASI ground station in Malindi (Kenya), the Telespazio Mission Control Center at Fucino, and the data processing done at the ASI/SSDC in Rome: the success of AGILE scientific operations depends on the effectiveness of the data flow from Kenya to SSDC and the data analysis and software management.

\bibliography{P414}  

\begin{thebibliography}{}
\expandafter\ifx\csname natexlab\endcsname\relax\def\natexlab#1{#1}\fi
\expandafter\ifx\csname url\endcsname\relax
  \def\url#1{\texttt{#1}}\fi
\expandafter\ifx\csname urlprefix\endcsname\relax\def\urlprefix{URL }\fi
\providecommand{\eprint}[2][]{\url{#2}}

\bibitem[{Acharya et~al.(2018)}]{2018}
Acharya, B.~S., et~al. 2018, Science with the Cherenkov Telescope Array (WORLD SCIENTIFIC). \urlprefix\url{http://dx.doi.org/10.1142/10986}

\bibitem[{Bergholm et~al.(2022)}]{bergholm2022pennylane}
Bergholm, V., et~al. 2022, Pennylane: Automatic differentiation of hybrid quantum-classical computations. \eprint{1811.04968}

\bibitem[{Broughton et~al.(2021)}]{broughton2021tensorflow}
Broughton, M., et~al. 2021, Tensorflow quantum: A software framework for quantum machine learning. \eprint{2003.02989}

\bibitem[{Bulgarelli et~al.(2010)}]{BULGARELLI2010213}
Bulgarelli, A., et~al. 2010, Nuclear Instruments and Methods in Physics Research Section A: Accelerators, Spectrometers, Detectors and Associated Equipment, 614, 213. \urlprefix\url{https://www.sciencedirect.com/science/article/pii/S0168900209023882}

\bibitem[{Hur et~al.(2022)Hur, Kim, \& Park}]{Hur_2022}
Hur, T., Kim, L., \& Park, D.~K. 2022, Quantum Machine Intelligence, 4. \urlprefix\url{http://dx.doi.org/10.1007/s42484-021-00061-x}

\bibitem[{Parmiggiani et~al.(2021)}]{Parmiggiani_2021}
Parmiggiani, N., et~al. 2021, The Astrophysical Journal, 914, 67. \urlprefix\url{https://doi.org/10.3847%2F1538-4357%2Fabfa15}

\bibitem[{Parmiggiani et~al.(2023)}]{Parmiggiani_2023}
--- 2023, The Astrophysical Journal, 945, 106. \urlprefix\url{https://dx.doi.org/10.3847/1538-4357/acba0a}

\bibitem[{Pérez-Salinas et~al.(2020)Pérez-Salinas, Cervera-Lierta, Gil-Fuster, \& Latorre}]{P_rez_Salinas_2020}
Pérez-Salinas, A., Cervera-Lierta, A., Gil-Fuster, E., \& Latorre, J.~I. 2020, Quantum, 4, 226. \urlprefix\url{http://dx.doi.org/10.22331/q-2020-02-06-226}

\bibitem[{{Qiskit contributors}(2023)}]{Qiskit}
{Qiskit contributors} 2023, Qiskit: An open-source framework for quantum computing

\bibitem[{{Tavani} et~al.(2008)}]{tavani2008}
{Tavani}, M., et~al. 2008, Nuclear Instruments and Methods in Physics Research A, 588, 52

\bibitem[{Tomsick et~al.(2019)}]{tomsick2019compton}
Tomsick, J.~A., et~al. 2019, The compton spectrometer and imager. \eprint{1908.04334}

\end{thebibliography}

\end{document}